 \documentstyle[12pt]{article}
\title{Morphological properties of short duration gamma ray bursts}
\author{Varsha Gupta\thanks{varsha@ducos.ernet.in}, \ Patrick Das Gupta\thanks{patrick@ducos.ernet.in} \\
{\it Department of Physics \& Astrophysics,} \\
{\it University of Delhi, Delhi 110 007, India.} \\
 and \\ 
P. N. Bhat\thanks{pnbhat@tifr.res.in}\\
{\it Tata Institute of Fundamental Research,} \\
{\it Homi Bhabha Road, Mumbai 400 005, India.}} 
\begin{document}
\maketitle
\begin{abstract}
 In this paper, we study a sample of 65 short duration bursts contained
 in the
 3B catalogue. We fit the time profiles of these GRBs with
 lognormal functions and study various  temporal properties.
 In most of the multi-peaked bursts, our analysis leads to
 a statistical evidence
 for the evolution of temporal asymmetry in individual pulses
 i.e. subsequent pulses in a GRB, on an average, tend to be
  more symmetric. Using
 a peak asymmetry evolution parameter, we find that $90 \%$  of 
short duration GRBs with 3 or more peaks exhibit the above trend
individually. 
\end{abstract}

 \section{Introduction}
 With the detection of afterglows in radio, optical and X-ray from
 several
 gamma ray burst events as well as with the measurement of redshifts
 corresponding to a few of
 these transient sources, there has been a tremendous 
 enhancement in the understanding
 of GRBs (e.g.\cite{met,wax,kul}). In particular, if the redshift estimates are correct,
 the extragalactic nature of GRBs stands confirmed, settling a three
 decade old
 controversy. However, as far as gamma emission is concerned, the
 diversity
 in time-profiles, spectral variability, duration etc. observed with
 regards
  GRBs still poses a major challenge to the theoretical models.
 For most GRBs, the gamma-ray time history shows complex structures
 with several
 peaks. A proper understanding of the origin of peaks is still
 lacking,
  although many ideas have been proposed in the literature (e.g.
 \cite{shav,feni,sari}).
  As a step to study such peaks in a systematic manner we direct our
  attention,in this paper, to 65 short duration
  GRBs ($T_{90}$ less than 2 seconds) contained in the 3B catalogue,
 and analyse their temporal properties
 after fitting the gamma ray time-history of each burst with a
 collection of   lognormal functions. 
 \section{Lognormal fits and data analysis}
 Individual peaks in most GRBs exhibit a rapid rise and slow decay,
 and
  therefore, it is convenient to fit each such peak with a lognormal
 function,
\begin{equation}
C(t)= \left\{ \begin{array}{ll}
    \frac{N}{t\sqrt{2\pi}\sigma}\exp{[{-(\log{t}-\mu)^{2}}/
               {2 {\sigma}^{2} }]} &  \mbox{ $t>0$} \\
                                0  &  \mbox{$t \leq 0$}
        \end{array}
        \right. 
\end{equation}

\noindent  where $C(t)$ is the photon counts in the peak at time $t$ while
 $N$, 
 $\mu $ and $\sigma $ are
 the associated free parameters that are fixed after obtaining a
 best fit.
  To begin with, the position of a peak is taken
 to be the point at which the first differential of the time history
 goes from 
 positive to negative. A low pass filter is applied several times to
 eliminate
 spurious peaks arising out of background noise.
  Having identified the genuine burst peaks, a reduced
 ${\chi_{\nu}}^2$
 is calculated after making an initial estimate  of free parameters 
 for
 each peak, and then these parameters are changed iteratively till
  ${\chi_{\nu}}^2$  changes by only $0.1$ percent. The final
  numerical values at which the iteration converges are taken to be
 the best-fit parameters. The fit is accepted only if
 the  final ${\chi_{\nu}}^2$ is close to $1$ (for details, see \cite{gup})
 and the best fit  parameters are then used to calculate the 
 following quantities associated with a given burst:
\begin{enumerate}
 \item Burst Complexity Index(CI): This is defined
 as the number of peaks detected in a given burst. We find that
 bursts 
 with higher CI are less in number, and that 
 average value  of CI  is  2.48  implying that on an average the
 bursts
 have 2 to 3 peaks. 
 \item Risetime $(t_{r})$ : Risetime of a peak is taken to be
 the
 time interval during which the value of the fitted lognormal
 function
 increases from $5 \%$  to $95 \%$ of its
 maximum value.  
 \item  Decaytime ($t_{d}$): Decaytime is the time interval
 during which the
 fitted counts in the peak 
 falls from $95\%$  to $5\%$ of its maximum value.
 \item $r_{rd}$ : $r_{rd}$ is defined as the ratio of $t_r$ to
 $t_d$. Since, most GRBs lack redshift information, use of
 this ratio in the analysis has a merit in the sense that because of
 the cancellation of stretching of $t_r$ and $t_d$ due
 to cosmological expansion, $r_{rd}$ is independent of burst
 redshift. 
 We compute this ratio for the first peak, second peak and so on for
 all bursts with a given CI, and then consider the average value of 
 $r_{rd}$ corresponding to first, second and so on. Since, $r_{rd}$
 does not depend on the distance of the GRB event, any systematic
 trend
 exhibited by the value of $r_{rd}$ averaged over different bursts
 with a given CI has a physical significance. The results have been
 summarized in Table 1. It is evident from the table that later
 peaks tend to
 be more symmetric than the earlier ones in a monotonic fashion.
 \item Peak Asymmetry Evolution Parameter (PAEP):
 In order to study the evolution of $r_{rd}$ within a burst, we plot
 $r_{rd}$ against the peak number for bursts with CI $\geq $ 3 (27
 of them) and fit with a straight line.
 PAEP is defined
 to
 be the slope of the fitted line - a positive value indicates that
 the
 peaks in the corresponding burst tend to become more symmetric as
 time evolves.
  In our analysis, we find that PAEP is positive in $24$ cases out
 of 27 bursts,
   while only 3 bursts
 show  negative values for PAEP. We provide a histogram of PAEP in
 Fig.1. The average PAEP observed for bursts with CI $\geq $ 3 is
 $.127 \pm .116$ while the histogram attains its maximum value when
 PAEP $\sim $ 0.1. 
\end{enumerate}
 \section{Conclusion}
 Short duration GRBs exhibit a strong evolutionary trend - later
 peaks tend to 
 assume a more symmetric time profile. This is seen both on an
 average 
 in bursts with any CI (as is apparent from table 1) as well as in
 individual
 bursts, evident from the fact that PAEP is positive in 90 $\% $  of
 the cases studied. 
 
\begin{section}*{ Acknowledgements}
One of us (VG) wishes to thank both Dr.Varsha Chitnis for useful
suggestions as well as Tata Institute of Fundamental Research, Mumbai,
for providing partial financial assistance and research facilities. VG
also takes this opportunity to thank University Grants Commission, New
Delhi, for the award of Senior Research Fellowship.
 \end{section}

\begin{table}
\begin{center}
\begin{tabular}{|l|lllll|}\hline
$Burst \ Sample $ & $ {<r_{rd}>}_1 $ & $ {<r_{rd}>}_2$ & $ {<r_{rd}>}_3$ & $ {<r_{rd}>}_4$ & $ {<r_{rd}>}_5$ \\   \hline
$CI =1$ & $.29 \pm .14$ & & & &\\ 
$CI =2 $ & $.46 \pm .24$ & $.57 \pm .27$ & & & \\
$CI=3$ & $.44 \pm .21$ & $.76 \pm .14$ & $.77 \pm .18$ & &  \\
$CI=4 $ & $.41 \pm .24$ & $.74 \pm .19$ & $.85 \pm .06$ & $.69 \pm .15$ &  \\
$CI=5$ & $.48 \pm .23$ & $.62 \pm .23$  & $.73 \pm .10$  & $.85 \pm .08$ & $.84\pm .12$ \\
\hline
\end{tabular}
\caption{ Value of $r_{rd}$ of subsequent pulses
 averaged over  bursts with same Complexity Index. $<r_{rd}>_i$ is the average
 $r_{rd}$ of $i^{th}$ peak in a burst, i=1 to 5.   }
\end{center}
\end{table}

\newpage
\begin {figure}[ht]
\vskip 15truecm
 \includegraphics{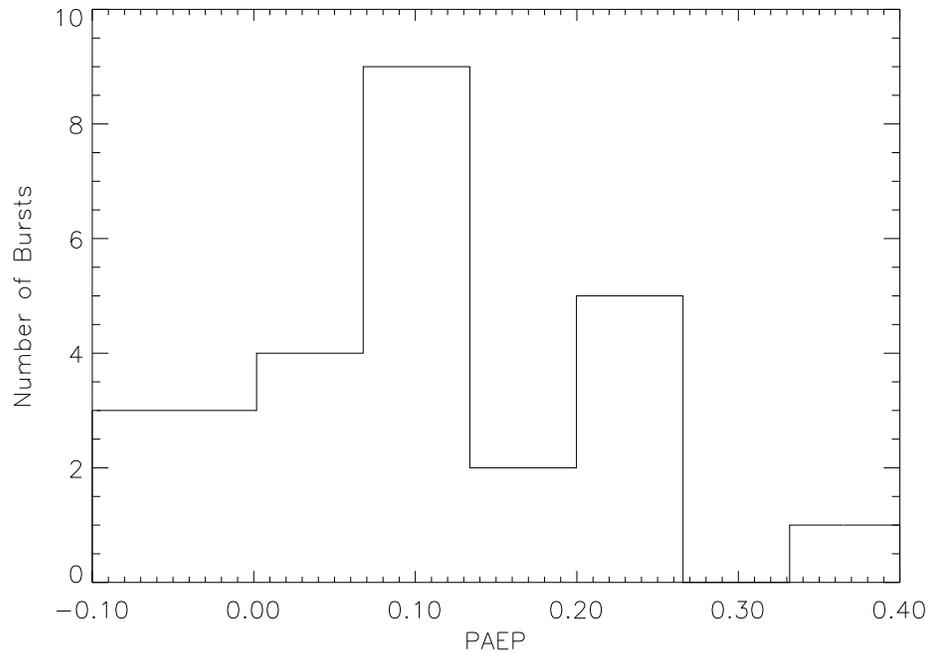}

\caption { Histogram of PAEP shows that 24 out of 27 bursts
          are associated with a positive PAEP,
          reinforcing the conclusion that peaks tend to become more
          symmetric
          with time. }
\end {figure} 
\end{document}